\documentclass[floatfix,pra,superscriptaddress,twocolumn]{revtex4}
\usepackage{optidef}

\usepackage{graphicx,bm,natbib,upgreek,amsmath,mathrsfs,accents}
\usepackage{amsbsy}
\usepackage{amsfonts}

\usepackage{amsthm}

\usepackage[dvipsnames]{xcolor}

\usepackage{graphicx}
\usepackage{subfigure}

\usepackage{hyperref}

\usepackage[nice]{nicefrac}

\usepackage[normalem]{ulem}

\usepackage{color}

\begin{document}

\title{Single-photon super-resolved spectroscopy from spatial-mode demultiplexing}

\author{Luigi Santamaria}
\author{Fabrizio Sgobba}
\author{Deborah Pallotti}
\affiliation{Agenzia Spaziale Italiana, Matera Space Center, Contrada Terlecchia snc.~75100 Matera, Italy}

\author{Cosmo Lupo}
\affiliation{Dipartimento Interateneo di Fisica, Politecnico \& Universit\`a di Bari, 70126 Bari, Italy}
\affiliation{INFN, Sezione di Bari, 70126 Bari, Italy}


\begin{abstract}
We demonstrate spectroscopy of incoherent light with sub-diffraction resolution. 
In a proof-of-principle experiment we analyze the spectrum of a pair of incoherent point-like sources whose separation is below the diffraction limit.
The two sources mimic a planetary system, with a brighter source for the star and a dimmer one for the planet.
Acquiring spectral information about the secondary source is hard because the two images have a substantial overlap.
This limitation is solved by leveraging a structured measurement based on spatial-mode demultiplexing, where light is first sorted in its Hermite-Gaussian components in the transverse field, then measured by photon detection. 
This allows us to effectively decouple the photons coming from the two sources.
An application is suggested to enhance exoplanets' atmosphere spectroscopy.
A number of experiments of super-resolution imaging based on spatial demultiplexing have been conducted in the past few years, with promising results. 
Here, for the first time to the best of our knowledge, we extend this concept to the domain of spectroscopy.   
\end{abstract}

\maketitle


\section{Introduction}

Establishing and achieving the ultimate limit of optical resolution are long-standing problems in physics with profound and widespread impacts on a number of disciplines, from engineering to medical practice.
Continuous advances in physics have the disruptive capability of redefining such limit pushing the boundaries of image resolution.

New imaging systems are commonly benchmarked against the well-known Rayleigh resolution criterion.
Due to the wave-like nature of light, the image of a point-source is a spot characterized by the point-spread function (PSF) of the optical system.
The Rayleigh criterion establishes that two point-sources are hard to resolve if their transverse separation is smaller than the width of the PSF, which in turn is determined by the Rayleigh length 
$\mathrm{x_R} = \lambda D/R$, where $\lambda$ is the wavelength, $R$ the radius of the pupil of the optical system, and $D$ the distance to the object~\cite{Rayleigh1879}.
Notable methodologies that bypass the Rayleigh resolution criterion include fluorescence microscopy, which exploits controlled activation and de-activation of neighboring emitters~\cite{Hell1994}; and quantum optics, which exploits the fact that states with exactly $N$ photons have an effective $N$-times smaller wavelength~\cite{JPD,Giovannetti}.

Inspired by quantum metrology, whose general goal is to exploit quantum physics to boost measurement devices and methodologies, a modern formulation of the Rayleigh criterion has been recently proposed~\cite{Modern} inspired by the seminal work by Tsang and collaborators~\cite{Tsang2016}.
The key observation is that the classic Rayleigh resolution criterion applies only to standard imaging approach where the intensity of the focused light is measured pixel-by-pixel on the image plane.
However, optics offers much more than direct intensity detection and a structured measurement may be able to extract the information carried by the phase of the light field.
In a typical structured measurement, light is first pre-processed, for example in a multiport interferometer, and then measured~\cite{gmachine,Sidhu2023linearoptics}. This accounts to sorting the light field into a suitably defined set of optical modes, followed by mode-wise photo detection.
Spatial-mode demultiplexing (SPADE) is a particularly efficient way of measuring the focused optical field. It has been shown to be optimal for a number of problems~\cite{Tsang2016,Modern,lu2018quantum,Dutton,GraceGuha,PhysRevLett.117.190802,PhysRevLett.117.190801,PhysRevLett.127.130502}, including the textbook problem of resolving two neighboring point-like sources.
In SPADE, the field is de-multiplexed in its transverse degrees of freedom. Oftentimes the transverse field is decomposed into the Hermite-Gaussian (HG) modes. Experimentally, SPADE can be implemented using a number of methodologies, including spatial light modulators~\cite{paur2016achieving,Paur:18,Zhou2019,Lvovsky,Zhou:23,Frank:23}, 
image inversion~\cite{Tham2017}, 
photonic lantern~\cite{Salit:20}, and multi-plane light conversion~\cite{Boucher2020,Santamaria22,Treps23,Amato23}.

Theoretical analyses show that SPADE is in principle the overall optimal measurement for a number of estimation and discrimination problems, with a notable improvement in performance with respect to direct detection (DD), both in the single-photon regime and for bright sources~\cite{Lvovsky,Boucher2020,Santamaria22,Treps23,Frank:23}. 
Practical implementations are affected by cross talk~\cite{
Gessner2020,Boucher2020,ct2022,Treps23,Schlichtholz:24}, which represent the most important limitation of this methodology. 


The natural field of application of SPADE as a super-resolution detection methodology is within astronomical imaging. Proposed use cases include the estimation of the angular separation between stars~\cite{Tsang2016}, or between a star and its exoplanet~\cite{PhysRevA.96.062107}, and the detection of the presence of an exoplanet in the vicinity of a star~\cite{PhysRevLett.127.130502,GraceGuha}.
All these applications refer to spatial degrees of freedom with the goal of extracting information about the spatial distribution of the source intensity in the transverse plane.
Going beyond spatial degrees for freedom, SPADE may also be exploited to enhance spectroscopy. 
In a recent work, some of us have proposed to use SPADE to boost exoplanet spectroscopy~\cite{astro22}, with a potential use to detect bio-signature on distant planets.
In this work we develop on this idea and present, for the first time to the best of our knowledge,  an experimental demonstration of super-resolved spectroscopy. 
The key enabler is the correlation between spatial and frequency degrees of freedom, e.g.~when the spectrum of the exoplanet is slightly different from the spectrum of the star due absorption lines. 
SPADE is useful as it helps us in sorting which photons are coming from the exoplanet and distinguish them from those coming directly from the star. These photons are hard to distinguish using direct detection, since the star is typically much brighter than the planet, and the two may have small angular separation with respect to the PSF of the optical system.

The paper develops as follows: 
in Section~\ref{sec:mode} we introduce the physical model and our assumptions.
In Section~\ref{sec:fisher} we present an analysis of exoplanet spectroscopy developed in terms of Fisher information and quantum Fisher information. This allows us to show how SPADE allows super-resolution spectroscopy.
Section~\ref{sec:setup} introduces our experimental setup and discusses how it can be used to simulate the observation of a planetary system.
The experiment, including calibration and measurement outcomes, is discussed in Section~\ref{sec:results}.
Finally, conclusions are presented in Section~\ref{sec:end}.


\section{Physical model and assumptions}\label{sec:mode}

Consider a point-like source located in position $x$ in the object plane that emits incoherent, monochromatic light at wavelength $\lambda$. 
Through an optical system, a focused image is produced in the image plane. 
In the far-field and paraxial approximation, the field in the image plane is characterized by the PSF of the optical system, denoted as $T$~\cite{goodman2008introduction}.
The width of the PSF is determined by the Rayleigh length, and its shape depends on the pupil of the optical system. As long as the source is much larger than the wavelength, it is sufficient to use a scalar theory for the light field. Therefore, the scalar field in position $y$ in the image field is proportional to $T(y/M-x)$, where $M$ is the magnification factor.

In linear optics, the classical scalar theory can be directly lifted into the quantum theory, and the PSF gives the rule of transformation of the field operators in the object plane into those in the image plane~\cite{Shapiro,PhysRevA.84.010303,PhysRevA.85.062314}.
Here we are interested in the single-photon regime, where at most one photon is detected in the image plane per detection window. It is therefore sufficient to consider a single photon emitted by the source, and how this photon arrives on the image plane. 
A single-photon state emitted at point $x$ in the object plane is described by the state
\begin{align}
| 1_{x,\lambda} \rangle = a^\dag_{x,\lambda} |0\rangle \, ,
\end{align}
where $\{ a_{x,\lambda}, a^\dag_{x,\lambda}\}$ are the canonical operators that annihilate and create a photon of wavelength $\lambda$ at position $x$ in the object plane.
When such a state passes through a diffraction-limited optical system, characterized by the PSF $T$, the single-photon state in the image plane reads
\begin{align}
| T_{x,\lambda} \rangle = \int dy \, T(y-x) b^\dag_{y,\lambda} |0\rangle \, ,
\end{align}
where $\{ b_{y,\lambda}, b^\dag_{y,\lambda}\}$ are the canonical operators that annihilate and create a photon of wavelength $\lambda$ at position $y$ in the image plane. Here to simplify the notation we have put $M=1$.

From a single point-like source we now move to the case of an extended incoherent source. Consider that a photon of wavelength $\lambda$ is emitted from position $x$ with probability $p(x,\lambda)$. Then the state of the single-photon in the image plane is represented by the density matrix
\begin{align}
    \rho = \sum_{x,\lambda} p(x,\lambda) | T_{x,\lambda} \rangle \langle T_{x,\lambda} | \, .
\end{align}

Our goal is to model the observation of light from a exoplanetary system, where the light coming from the star (the primary source) is also scattered by the exoplanet (secondary source). Due to absorption through the atmosphere of the planet, the two sources may have different spectra.
Modeling star and planet as point-like sources, the state of a single photon collected in the image plane is
\begin{align}\label{state0}
\rho & = (1-\epsilon) \sum_\lambda f_s(\lambda) 
|T_{x_s,\lambda}\rangle \langle T_{x_s,\lambda} |
\nonumber \\
& \phantom{=}~+
\epsilon \sum_\lambda f_p(\lambda) |T_{x_p,\lambda}\rangle \langle T_{x_p,\lambda}| \, .
\end{align}
Here, $\epsilon$ and $(1-\epsilon)$ are the relative intensities of planet and star, $|T_{x_\star,\lambda}\rangle$ represents the state of a single photon at wavelength $\lambda$, emitted either from the star ($\star = s$) or planet ($\star = p$), and $f_\star(\lambda)$ is the corresponding spectrum.
The objective of exoplanet spectroscopy is to obtain information about the spectrum $f_p(\lambda)$ of the planet. This task comes with at least two challenges: 
(1) The star is much brighter than the planet, therefore most of the photons detected come from the star;
(2) If the transverse separation between star and planet is below the Rayleigh length, then the PSFs $T_{x_p,\lambda}$ and $T_{x_s,\lambda}$ will overlap substantially, making hard to distinguish which photons are coming from the planet.

To make our presentation more concrete, we assume a Gaussian PSF. 
Formally, this would follow from a pupil function with a Gaussian profile. However, a Gaussian PSF is also a good approximation for other PSFs with enough regularity~\cite{Santamaria22} in the sub-diffraction regime where the object is much smaller than the width of the PSF: this is exactly the regime we are interested in. A Gaussian PSF reads
\begin{align}
    T(y-x) = \mathcal{N} \, e^{-\frac{(y-x)^2}{4\sigma^2}} \, ,
\end{align}
where the factor $\mathcal{N}$ is to normalize the PSF to one, i.e., $\int dy |T(y-x)|^2 =1$.
The width of the Gaussian can be identified with the Rayleigh length, $\sigma \equiv \mathrm{x_R}$. It follows that in general the PSF depends on the wavelength through its width. Here we assume a regime where the observed spectral range is narrow enough that the width of the PSF may be assumed constant.

Direct detection of the state in Eq.~(\ref{state0}) amounts to address the field in each pixel in the image plane, and to apply a spectral analyzer pixel by pixel.
The outcome of such a measurement strategy is described by the probability of detecting a photon at position $y$ with wavelength $\lambda$, \begin{align}
p_\text{DD}(y,\lambda) 
& = (1-\epsilon) f_s(\lambda) 
|\langle 1_{y,\lambda}| T_{x_s, \lambda} \rangle|^2
\nonumber \\
& \phantom{=}~+
\epsilon f_p(\lambda) 
|\langle 1_{y,\lambda}| T_{x_p, \lambda}\rangle|^2 
\\
& = (1-\epsilon) f_s(\lambda) \mathcal{N}^2 e^{ - (y-x_s)^2/2\sigma^2} \nonumber \\
& \phantom{=}~+
\epsilon f_p(\lambda) 
\mathcal{N}^2 e^{ - (y-x_p)^2/2\sigma^2} 
\, .
\label{pDD1}
\end{align}

We will compare direct detection with a methodology to obtain spectral measurements based on HG SPADE. 
In a structured measurement of the optical field, we first sort the transverse field in the image plane along basis elements $\{ |\Psi_u\rangle \}_{u-0,1,2,\dots}$.
Then, we apply a spectral analyzer to each mode. The outcomes of this measurement are described by the probability of measuring a photon of wavelength $\lambda$ in mode $u$:
\begin{align}
p_\text{SPADE}(u,\lambda) 
& = 
(1-\epsilon) f_s(\lambda) 
\left| \langle \Psi_u|T_{x_s,\lambda} \rangle \right|^2
\nonumber \\
& \phantom{=}~+
\epsilon f_p(\lambda) 
\left| \langle \Psi_u| T_{x_p,\lambda}\rangle \right|^2
\,.
\end{align}
HG SPADE is obtained when the basis elements are Hermite-Gaussian modes, i.e.,
\begin{align}
| \Psi_u \rangle = \int dy \, \Psi_u(y) b^\dag_{y,\lambda} |0\rangle \, ,
\end{align}
with
\begin{align}
\Psi_u(y) = \text{HG}_u(y)
=  
\frac{\mathcal{N} }{\sqrt{2^u u!}} \,
e^{ - \frac{ y^2 }{4 \sigma^2} }
H_u\left( \frac{y}{\sqrt{2}\sigma} \right) 
\, ,
\end{align}
and $H_u$ denotes the Hermite polynomials.
Note that the width $\sigma$ is chosen to match that of the Gaussian PSF.


\section{Analysis in terms of Fisher information}\label{sec:fisher}

To study the problem of spectral analysis with sub-diffraction resolution, and compare different detection strategy, we may use the Fisher information as a figure of merit.
Spectral analysis can be understood as a problem of multi-parameter estimation, whose goal is to estimate the spectrum of the exoplanet, $f_p(\lambda)$ for a range of discrete values of the wavelength $\lambda_1$, $\lambda_2$, $\dots$, $\lambda_N$.

For given $n$ photons detected, the uncertainty in the estimation of the spectrum due to statistical fluctuations is quantified by the covariance matrix $\Sigma$, with matrix elements
\begin{align}
   \Sigma_{ij} = \mathbb{E}\left[ f_p(\lambda_i) f_p(\lambda_j) \right]
   - \mathbb{E}\left[ f_p(\lambda_i) \right]
   \mathbb{E}\left[ f_p(\lambda_j) \right] \, ,
\end{align}
where $\mathbb{E}$ denotes the expectation value.
According to the Cramér-Rao bound, for a given measurement, the covariance matrix is lower bounded by the Fisher information matrix:
\begin{align}
   \Sigma \geq \frac{1}{n} F^{-1} \, ,
\end{align}
where $F^{-1}$ denotes the inverse matrix of the Fisher information matrix. For a given measurement described by the probability mass distribution $p(x)$, we have
\begin{align}\label{defF}
   F_{ij} & = \sum_x p(x) \frac{\partial \log{p(x)} }{\partial f_p(\lambda_i) }
   \frac{\partial \log{p(x)} }{\partial f_p(\lambda_j) }
   \, .
\end{align}


\subsection{Spectroscopy by direct detection}\label{sec:specDD}

In direct detection we have $x \equiv (y,\lambda)$.
From Eq.~(\ref{pDD1}) we obtain
\begin{align}
\frac{\partial \log{p_\text{DD}(y,\lambda)} }{\partial f_p(\lambda_i) }    
= \delta_{\lambda, \lambda_i}
\epsilon 
\mathcal{N}^2 e^{ - (y-x_p)^2/2\sigma^2}
\, , 
\end{align}
where $\delta_{\lambda, \lambda_i}$ is the Kronecker delta function.
Because to the presence of the $\delta_{\lambda, \lambda_i}$, the Fisher information matrix is diagonal, with elements
\begin{align}
    F^\text{DD}_{jj}
    & = 
    \sum_{y} 
p_\text{DD}(y,\lambda_j)
\left(  \frac{1}{p_\text{DD}(y,\lambda_j)}  \frac{\partial p_\text{DD}(y,\lambda_j) }{\partial f_p(\lambda_j) }
\right)^2
\, .
\end{align}
For a Gaussian PSF this reads
\begin{widetext}
\begin{align}
F^\text{DD}_{jj} 
    & = \sum_{y} p_\text{DD}(y,\lambda_j) \left( 
    \frac{ \epsilon \mathcal{N}^2 e^{ - (y-x_p)^2/2\sigma^2} }{p_\text{DD}(y,\lambda_j)}
    \right)^2 \\
    & = \sum_y p_\text{DD}(y,\lambda_j) \left( 
    \frac{ \epsilon e^{ - (y-x_p)^2/2\sigma^2} }{(1-\epsilon) f_s(\lambda_j) e^{ - (y-x_s)^2/2\sigma^2}
+
\epsilon f_p(\lambda_j) 
e^{ - (y-x_p)^2/2\sigma^2}}
    \right)^2 \\
& = \sum_y  
    \frac{ \epsilon^2 e^{ - (y-x_p)^2/\sigma^2} }{(1-\epsilon) f_s(\lambda_j) e^{ - (y-x_s)^2/2\sigma^2}
+
\epsilon f_p(\lambda_j) 
e^{ - (y-x_p)^2/2\sigma^2}} \, .
\label{eq17}
\end{align}
\end{widetext}

In the following, to make the notation lighter we put $F^\text{DD}(\lambda_j) := F^\text{DD}_{jj}$.

Note that the Fisher information can be interpreted as the average of the squared signal-to-noise ratio (SNR). 
The maximum value of the SNR is when $y=x_p$, yielding
\begin{align}
    F^\text{DD}(\lambda)
    & \leq \left( 
    \frac{ \epsilon }{(1-\epsilon) f_s(\lambda) e^{ - (x_p-x_s)^2/2\sigma^2}
+
\epsilon f_p(\lambda) 
}
    \right)^2 
    \label{DDscale}
\end{align}
(this follows from the fact the average is always smaller than the maximum).
This lower bound shows that the Fisher information from direct detection is of order $\epsilon^2$ unless the sources are well separated, i.e.
\begin{align}
     e^{ - (x_p-x_s)^2/2\sigma^2}
\ll \frac{\epsilon}{1-\epsilon} \frac{f_p(\lambda)}{f_s(\lambda)} \, .
\end{align}
If the source overlap substantially, then $F^\text{DD} \sim  \epsilon^2$. 
Otherwise, if they are well separated, then $F^\text{DD} \sim \epsilon$. In fact, from Eq.~(\ref{eq17}) we obtain
\begin{align}
F^\text{DD}(\lambda)
& \geq 
\frac{ \epsilon^2 }{(1-\epsilon) f_s(\lambda) e^{ - (x_p-x_s)^2/2\sigma^2}
+
\epsilon f_p(\lambda) 
}
\simeq 
\frac{ \epsilon }{
f_p(\lambda) 
}
\, .
\end{align}
(Here we have used the fact that the sum is always larger that one of the addends).


\subsection{Spectroscopy aided by HG SPADE}\label{sec:specHG}

The form of the state in Eq.~(\ref{state0}) implies that the Fisher information matrix is diagonal for any measurement. (In fact, the argument of Section~\ref{sec:specDD} applies to any measurement.)
For HG SPADE, we have
\begin{widetext}
\begin{align}
    F^\text{HG}(\lambda) 
    & = \sum_{u} p(u,\lambda)  
    \left( \frac{1}{p(u,\lambda)} \frac{\partial p(u,\lambda)}{\partial f_p(\lambda)} \right)^2 \\
    & = \sum_{u} p(u,\lambda)  
    \left( \frac{\epsilon |\langle \Psi_u | T_{x_p, \lambda} \rangle |^2 }{ (1-\epsilon) f_s(\lambda) \left| \langle \Psi_u|T_{x_s, \lambda}\rangle \right|^2
+
\epsilon f_p(\lambda) \left| \langle \Psi_u| T_{x_p, \lambda}\rangle \right|^2 }  \right)^2 
\, .
\end{align}
\end{widetext}

If the optical system is carefully aligned towards the star, the PSF matching the lower HG modes, most of the photons from the star will couple into the fundamental mode $\Psi_0$.
Also, in principle the photons from the star would not couple to mode $\Psi_1$ as the latter is an odd function with respect to the position of the star.
Finally, higher modes may be neglected in first approximation as they are highly suppressed in the sub-Rayleigh regime~\cite{GoodHelp}.
Overall, by considering only the lower HG modes, we have
\begin{widetext}
\begin{align}
    F^\text{HG}(\lambda) 
    & \simeq 
    p(0,\lambda)  
    \left( \frac{\epsilon |\langle \Psi_0 | T_{x_p, \lambda} \rangle |^2 }{ (1-\epsilon) f_s(\lambda) \left| \langle \Psi_0|T_{x_s, \lambda}\rangle \right|^2
+
\epsilon f_p(\lambda) \left| \langle \Psi_0| T_{x_p, \lambda}\rangle \right|^2 }  \right)^2 
\nonumber \\
& \phantom{=}~+
p(1,\lambda)  
    \left( \frac{\epsilon |\langle \Psi_1 | T_{x_p, \lambda} \rangle |^2 }{ (1-\epsilon) f_s(\lambda) \left| \langle \Psi_1|T_{x_s, \lambda}\rangle \right|^2
+
\epsilon f_p(\lambda) \left| \langle \Psi_1| T_{x_p, \lambda}\rangle \right|^2 }  \right)^2 
\label{HGfull}
\\
& \simeq 
    \frac{\epsilon^2 |\langle \Psi_0 | T_{x_p, \lambda} \rangle |^4 }{ (1-\epsilon) f_s(\lambda) \left| \langle \Psi_0|T_{x_s, \lambda}\rangle \right|^2
+
\epsilon f_p(\lambda) \left| \langle \Psi_0| T_{x_p, \lambda}\rangle \right|^2 }  
+
\frac{\epsilon |\langle \Psi_1 | T_{x_p, \lambda} \rangle |^2 }{ 
f_p(\lambda)  } 
\label{HGexpa} \\
& \geq 
\frac{\epsilon |\langle \Psi_1 | T_{x_p, \lambda} \rangle |^2 }{ 
f_p(\lambda)  }  \, .
\label{HGscale}
\end{align}
\end{widetext}

Note that the first term in (\ref{HGexpa}), which comes from photon detection in the fundamental mode $\Psi_0$, is proportional to $\epsilon^2$, whereas the second term, which come from detection in mode $\Psi_1$, is proportional to $\epsilon$.
The latter dominates in the regime where the planet is much dimmer than the star.

Comparing Eq.~(\ref{HGscale}) to Eq.~(\ref{DDscale}), we see that HG SPADE offers in principle a quadratic improvement in the scaling of the Fisher information when the transverse separation between star and planet is below the Rayleigh length. In the limit of ultra-weak signal coming from the planet, with direct detection the Fisher information vanishes with $\epsilon^2$, whereas the scaling is with $\epsilon$ in the case of HG SPADE.

By inspection of Eqs.~(\ref{HGfull})-(\ref{HGscale}), we note that the $O(\epsilon)$ scaling of the Fisher information crucially follows from the fact that 
there is a vanishing probability that a photon emitted from the star is detected in the mode $\Psi_1$, i.e., $\langle \Psi_1|T_{x_s, \lambda}\rangle = 0$ in Eq.~(\ref{HGfull}).
In practice, this condition cannot be matched exactly due to misalignment and cross talk~\cite{Gessner2020,Boucher2020,ct2022,Schlichtholz:24}. In both case, we will have a non-zero probability that a photon emitted from the star is detected in the mode $\Psi_1$. This implies that the term $\langle \Psi_1|T_{x_s, \lambda}\rangle$ does not vanish and the Fisher information will scale as $\epsilon^2$.

To quantify the deviation from the ideal case in HG SPADE, we may consider a fidelity functional between the planet spectrum $f_p(\lambda)$ and the signal $p(1,\lambda)$ detected in mode $\Psi_1$,
\begin{align}
    \mathcal{F} =    
    \sum_\lambda f_p(\lambda) p(1,\lambda) 
    \, .
\end{align}
When the cross talk is negligible, we have
\begin{align}
    \mathcal{F} \simeq 
    \epsilon \left| \langle \Psi_1 | T_{x_p,\lambda}\rangle \right|^2
    \, .
\end{align}
Otherwise, the fidelity decreases in the presence of cross talk.

In our experimental setup, we indeed have that the two spectra have a small overlap and they are nearly orthogonal. 
In this case, it makes sense to consider the scalar product between the probability vectors $p(0,\lambda)$ and $p(1,\lambda)$, which represents the probabilities of detection in modes $\Psi_0$ and $\Psi_1$.
We define
\begin{align}\label{SPdef0}
    \mathcal{S} =    
    \sum_\lambda p(0,\lambda) p(1,\lambda) 
    \, .
\end{align}
Assuming non-overlapping spectra and negligible cross talk, this quantity is close to zero, $\mathcal{S} \simeq 0$. Otherwise, it increases with increasing cross talk, in principle up to $\mathcal{S}=1$ when cross talk is so intense that the signal is equally spread on both modes.


\subsection{Ultimate quantum limit of spectroscopy}

In general, the Fisher information depends on the chosen measurement strategy. The global bound on the covariance matrix for all possible measurements allowed by the principles of quantum mechanics is given by the quantum Cramér-Rao bound, 
\begin{align}
   \Sigma \geq \frac{1}{n} Q^{-1} \, ,
\end{align}
where $Q$ denotes the quantum Fisher information matrix.
Due to the form of the state in Eq.~(\ref{state0}), the quantum Fisher information is necessarily diagonal.
This follows from the fact that photonic states at different wavelengths are mutually orthogonal (this is an extension to the quantum domain of the argument in Sections~\ref{sec:specDD}).
The elements of $Q$ corresponding to the estimation of $f_p(\lambda)$ can be computed from the results of Ref.~\cite{Amato23} (which in turn are derived from~\cite{PhysRevA.96.062107}) after a simple change of variables. We obtain
\begin{align}
    Q(\lambda) = \frac{\epsilon}{1-\epsilon} \frac{1-w}{f_p(\lambda) } \, ,
\end{align}
where $w$ is a non-negative quantity~\cite{PhysRevA.96.062107}.
This expression matches Eq.~(\ref{HGscale}) showing the optimality of the linear  scaling of the Fisher information with $\epsilon$.


\section{Experimental setup}\label{sec:setup}

\begin{figure}
\centering
\includegraphics[scale=.27]{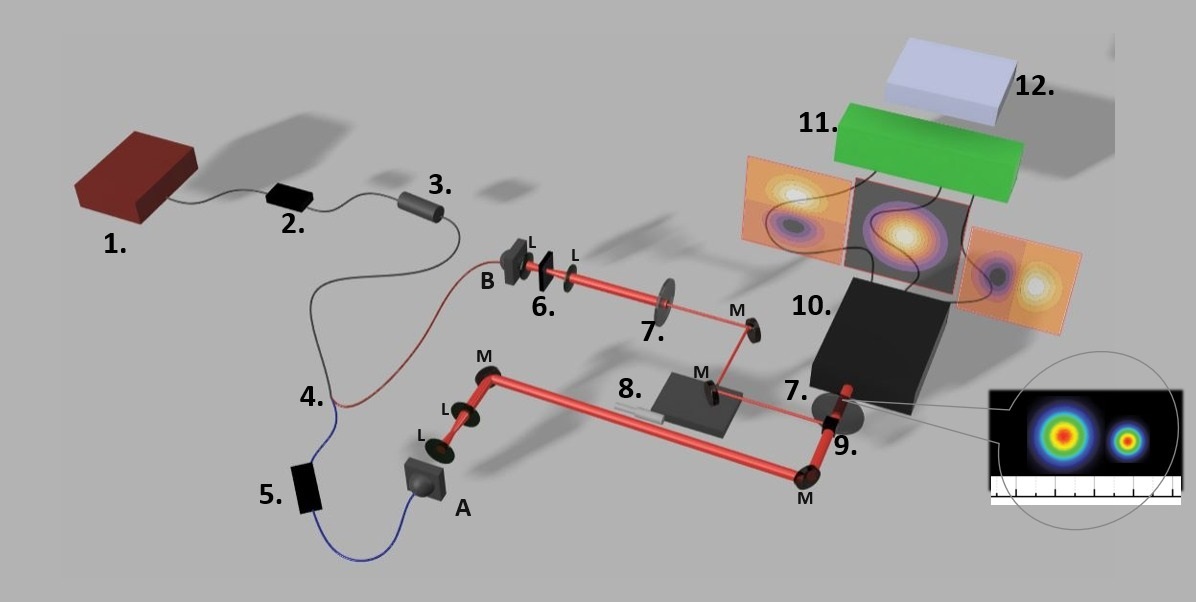}
\caption{\textbf{Experimental setup.}  A fiber coupled light emitting diode (LED) at telecom wavelength (1) inputs a motorized spectral filter (2), a 50 dB attenuator (3) and is split by a fiber beam splitter (4) to generate strong A beam simulating the star and weak B beam simulating exoplanet.
The A beam is carved from LED using a fiber coupled filter (5) and free space launched with collimator (A); the B beam  is free space launched with collimator (B) and carved using the free space spectral filter (6).
Both beams cross two-lens (L) systems to match the beams waist with demultiplexer waist ($300$ $\mu m$) and are coupled with it by means of two steering mirrors each. 
The B beam crosses a film polarizer mounted on a motorized rotation stage (7) to control the beam intensity.
The second mirror of B beam is placed on a micrometer translation stage (8) to shift the beam position and so change the A - B separation in controlled way.
A and B beams are recombined on a beam splitter (9) and, after crossing a fixed film polarizer (7'), are coupled with demultiplexer. This polarizer is used to keep fixed the photons polarization during the experiment to prevent polarization dependence of detection efficiency.
The demultiplexer (10), PROTEUS-C from Cailabs, allows to perform intensity measurements on six HG mode, but just the $\text{HG}_{01}$,  $\text{HG}_{10}$ and $\text{HG}_{00}$ modes are detected through three superconductive nanowires single photon detectors (11) whose electrical output signal is detected and counted by a time tagger (12). Three polarisation-rotation paddles for detectors efficiency maximisation are inserted between demultiplexer fiber outputs and nanowires fibers inputs and optimized at the beginning of the experiment.}
\label{fi:setup}
\end{figure}

\begin{figure}
\centering
\includegraphics[scale=.45]{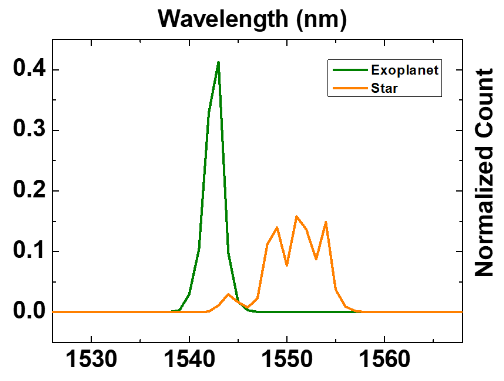}
\caption{\textbf{Point-like sources spectra.} The figure shows the spectra, obtained by normalizing single photon counts, of two point like sources simulating the star (orange line) and exoplanet (green line). The scan over the wavelength is performed using motorized fiber coupled spectral filter with 1 nm as minimum settable step size.}
\label{fi:Spectra}
\end{figure}

In a proof-of-principle experiment, we simulate the use of HG SPADE to achieve super-resolved spectroscopy of a planetary system.
The experimental setup is shown in Fig.~\ref{fi:setup}. 
We use a fiber-coupled Light Emitting Diode (LED) at telecom wavelength to simulate two incoherent point-like emitters. 
The LED is attenuated using fiber attenuators and, by means of a fiber beam splitter (FBS), it is split into two fiber-coupled beams:
the beam A simulates the star and the beam B simulates the exoplanet. 
Beam A is filtered using a fiber-coupled filter to induce a spectral shape different from that of beam B (described below), as shown in Fig.~\ref{fi:Spectra}. 
Then it is free-space collimated using a fiber collimator and a two-lens system to obtain a beam waist $w_0 \simeq 300 \, \mu m$. 
The A beam, through two steering mirrors, impinges on a cube beam splitter (BS) to be combined with the B beam and, after crossing a fixed-film polarizer, is coupled with the free-space input port of a $300 \, \mu m$ waist HG demultiplexer (PROTEUS-C model from Cailabs).

The other output of the FBS (B beam) is collimated using a fiber collimator and a two-lens system to obtain a beam waist $w_0 \simeq 300 \, \mu m$. 
Then it crosses a free-space spectral filter generating the spectral shape of the simulated exoplanet (shown in Fig.~\ref{fi:Spectra}).
The B beam crosses a polarizer mounted on a motorized rotation stage and two steering mirrors, one of which is mounted on a micrometer translation stage. 
Then it is reflected from the cube BS, where is combined with A beam, and crosses the fixed polarizer to overcome the dependency of detector efficiency on photon polarization when the stage is rotated. 
Finally the B beam is coupled with the HG demultiplexer.
In this way, by translating the translation stage, is possible to move the B beam and change the separation $d_a$ between the two beams. 
Moreover, by rotating the rotation stage the B beam intensity is tuned and the intensity ratio $\epsilon = N_B/N_A$ changes, where $N_A$ and $N_B$ are the number of photons impinging on demultiplexer emitted by A and B sources respectively.
The position and the intensity of the beam A remains fixed during the experiment.
Once the two beams entered the demultiplexer, they are decomposed in the lowest-order HG modes:
$\text{HG}_{00}$, $\text{HG}_{10}$, $\text{HG}_{01}$,  
$\text{HG}_{11}$, $\text{HG}_{20}$, $\text{HG}_{02}$. 
The modes are converted in $\text{HG}_{00}$ mode again and coupled with six single-mode fibers.
Finally, the modes 
$\text{HG}_{00}$, $\text{HG}_{01}$ and 
$\text{HG}_{10}$ are coupled to He-cooled superconducting nanowires single-photon detectors (NSPD) through single-mode fibers equipped with polarization paddles to rotate the photon polarization and maximize the NSPDs efficiency.

When photons impinge on the nanowires, the detectors generate electric pulses that, if exceeding the setup threshold, are recorded and counted by a commercial time tagger in the set temporal window of $50 \, ms$. The overall detectors quantum efficiency (system efficiency) is about $80$ per cent, the dark count is lower than $20 \, Hz$ and reset time is about $100 \, ns$.

When the beams completely overlap (the separation $d_a$ between simulated point like emitters reduces zero), the overall A+B beam symmetry is circular and the power leaked into first order modes 
($\text{HG}_{01}$ and $\text{HG}_{10}$) 
is minimum. 
Such a minimum value, named cross talk, is due to demultiplexer manufacturing imperfections.

In general, the cross talk between $\text{HG}_{00}$ and $\text{HG}_{nm}$ is defined by the ratio 
$P_{nm} / P_{00}$, 
where $P_{nm}$ is the power on the 
$\text{HG}_{nm}$ output fiber when only 
$\text{HG}_{00}$ is injected with a $P_{00}$ power in the input.
The cross talk is the main limiting factor in these kind of experiments and is due to some signal ending up in high-order modes even if the incoming light is fully matched, in term of waist and direction, with the fundamental mode (if the crosstalk were negligible only $\text{HG}_{00}$ should be excited for a fully matched incoming radiation). 
The measured cross talk $\chi$ of the first order modes in our case is $\chi = P_{01} / P_{00} + P_{10} / P_{00} \simeq 0.0035$.

\section{Calibration and Measurements}\label{sec:results}

The aim of the experimental work is to understand how HG SPADE can be used to perform spectroscopy of B source (representing the exoplanet) when a strong A source (representing the star) is at a separation lower than Rayleigh distance. 
For this purpose a motorized tunable filter (MTF) is inserted at the output of the LED (it is equivalent to put it at the input of the demultiplexer). 
It selects a given wavelength window (FWHM = $1 \, nm$) and is scanned between $1520 \, nm$ and $1569 \, nm$. 
In this way the system is able to acquire the spectrum of $\text{HG}_{00}$, $\text{HG}_{10}$ and $\text{HG}_{01}$.

In the experiment we operate in the single-photon regime and align the optical system to the brighter source. 
This is in contrast to what is done in other works, where the alignment is on the centroid or median point~\cite{Tsang2016,Nair:16,Paur:18,paur2016achieving,GraceGuha,Treps23}.
In our case, as the A source is much brighter than the B source, the procedure essentially coincides with alignment with the centroid, see Ref.~\cite{Amato23} for more details.
We want to stress that in case of a real observation the only way to proceed is by aligning the optical system to the centroid, as the median point is unknown and the sources may have an arbitrary shape.

To align the demultiplexer, we superimposed the B beam with the A beam (by setting the translation stage at $0 \, \mu m$) and maximize the signal in the $\text{HG}_{00}$ mode.
We collected the counts $C_0(\lambda)$ and $C_1(\lambda)$ in fundamental ($\text{HG}_{00}$) and first order modes ($\text{HG}_1\equiv \text{HG}_{01}+\text{HG}_{10}$) as function of wavelength $\lambda$ and for different values of the (dimensionless) source separation $d_a = d/w_0$ and intensity ratio $\epsilon = N_B/N_A$.
The counts are then normalized, obtaining 
\begin{align}
C_0^N (\lambda) & = \frac{ C_0(\lambda) }{\sqrt{ \sum_{\lambda'} C_0(\lambda')^2 }} \, , 
\label{CNeq1} \\
C_1^N(\lambda) & = \frac{C_1(\lambda)}{ \sqrt{ \sum_{\lambda'} C_1(\lambda')^2}} \, ,
\label{CNeq2}
\end{align}
which can be interpreted as a pair of vectors with 50 elements (one for each wavelength). 
In the experiment the total number of photons $N_A$ impinging on demultiplexer is fixed,
\begin{align}
N_A = 
\int_{1520 \, nm}^{1569 \, nm} 
N_A(\lambda) \, d\lambda \simeq 148000   
\end{align}
during the scanning time $t_s \simeq 2.5 s$ ($50 \, ms$ time windows for each of the 50 wavelengths).

\begin{figure}
\centering
\includegraphics[scale=.45]{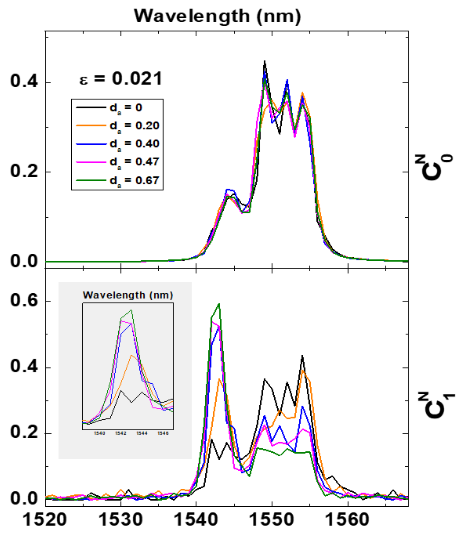}
\caption{\textbf{HG demultiplexed spectra at fixed $\epsilon$.} The figure shows the spectra of zero order (upper) and first orders (lower) HG modes obtained by normalizing the counts $C_0^N$ and $C_1^N$ respectively, as described in the text, at five values of separation $d_a$ and fixed $\epsilon = 0.021$.
The $C_1^N$ spectra show a strong dependency on $d_a$ proving the effectiveness of HG SPADE in spectroscopy of sources emitting spatial (without circular symmetry) dependent spectra. The inset of lower panel shows a magnification of the wavelength region due to simulated exoplanet emission where the dependency is even more clear.}
\label{fi:davar}
\end{figure}

Figure~\ref{fi:davar} shows the counts $C_0^N(\lambda)$ (upper panel) and  $C_1^N(\lambda)$ (lower panel) as functions of wavelength for five values of separation $d_a$. 
This figure shows that spectral information from the exoplanet is enhanced, in comparison to that from the star, when observing the mode $\text{HG}_{1}$.
The spectra $C_0^N(\lambda)$ are nearly independent on the separation, unlike $C_1^N(\lambda)$.  
This corresponds to the fact that the background from the star is mostly coupled into the fundamental mode $\text{HG}_{00}$.
In particular the left parts of the spectra (where the emission of the exoplanet is centered) are strongly dependent on separation $d_a$ as highlighted in the inset where the portion of the spectrum between $1538 \, nm$ and $1547 \, nm$ is shown.
The spectra appear quite rough because of filter minimum step ($1 \, nm$), but in principle it would be possible to obtain higher spectral resolution by using a finer step or a grating dispersing light on an array of single photon detectors.


\begin{figure}
\centering
\includegraphics[scale=.45]{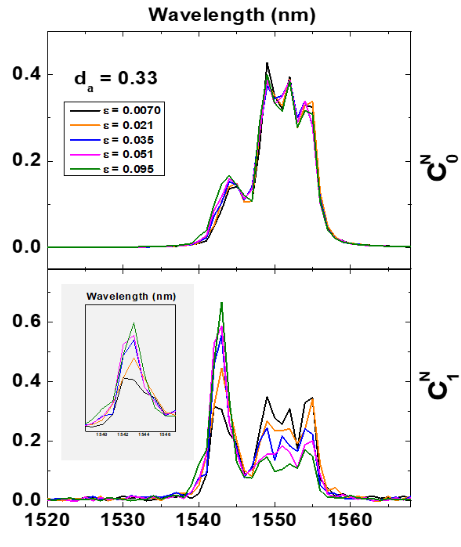}
\caption{\textbf{HG demultiplexed Spectra at fixed $d_a$.} The figure shows the spectra of zero order (upper) and first orders (lower) HG modes obtained by normalizing the counts $C_0^N$ and $C_1^N$ respectively, as described in the text, at five values of sources intensity ratio \textbf{$\epsilon = 0.021$}  and fixed separation $d_a$.
The $C_1^N$ spectra show a strong dependency on $\epsilon = 0.021$ proving the effectiveness of HG SPADE in spectroscopy of sources emitting spatial (without circular symmetry) dependent spectra. The inset of lower panel shows a magnification of the wavelength region due to simulated exoplanet emission where the dependency is even more clear.}
\label{fi:epsvar}
\end{figure}

Similarly, Fig.~\ref{fi:epsvar} shows the counts $C_0^N(\lambda)$ (upper panel) and  $C_1^N(\lambda)$ (lower panel) as functions of wavelength for five values of intensity ratio $\epsilon$. 
As above, the spectra $C_0^N(\lambda)$ are nearly independent on $\epsilon$, whereas $C_1^N(\lambda)$ are not. 
Also, the left parts of the spectra are strongly dependent on $\epsilon$, as highlighted in the inset.

To quantify the distinguishability of the photons coming from the simulated exoplanet in terms of their spectral properties, we may consider the scalar product between the vectors $C_0^N(\lambda)$ and $C_1^N(\lambda)$:
\begin{align}\label{Spdef}
\mathcal{S}_P = \sum_{\lambda}
C_0^N(\lambda) C_1^N(\lambda) \, .
\end{align}
Up to a different choice of normalization, this quantity is equivalent to the scalar product introduced in Eq.~(\ref{SPdef0}).
Since in our setup the spectra of planet and star are nearly orthogonal (they have nearly non-overlapping support in the spectral domain), this scalar product is close to zero when the system is able to distinguish the photons coming from the exoplanet from those coming directly from the star. 
Otherwise, as discussed in Section \ref{sec:fisher}, cross talk may increase the value of $\mathcal{S}_P$ up to the point where the photons becomes hardly distinguishable and $\mathcal{S}_P \simeq 1$.

\begin{figure}
\centering
\includegraphics[scale=.45]{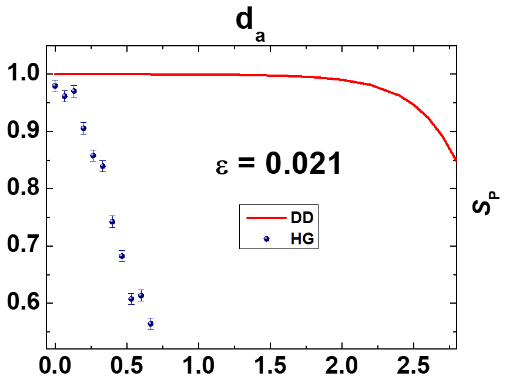}
\caption{\textbf{Scalar product quantifier.} 
The data points show the experimentally estimated scalar product defined in Eq.~(\ref{Spdef}), plotted as function of $d_a$ at fixed $\epsilon$; 
$\mathcal{S}_P(d_a,\epsilon=0.021)$
The error bars are estimated by repeated measure and take only into account statistical errors.
At $d_a=0$ the scalar product is slightly lower than unity (about 0.98) because of systematic uncertainty is not considered in error bars. By considering both statistic and systematic uncertainties we have an overall uncertainty of about $0.03$ and so, we can affirm, that when HG scalar product is below $0.97$ (for $d_a$ larger than $0.2$), the SPADE technique can distinctly discriminate between star and exoplanet.
The solid line shows the analogous quantity computed in the case of direct imaging.
In the case of DD a comparable discrimination capability is achieved when $d_a$ is larger than two.}
\label{fi:spd}
\end{figure}

\begin{figure}
\centering
\includegraphics[scale=.45]{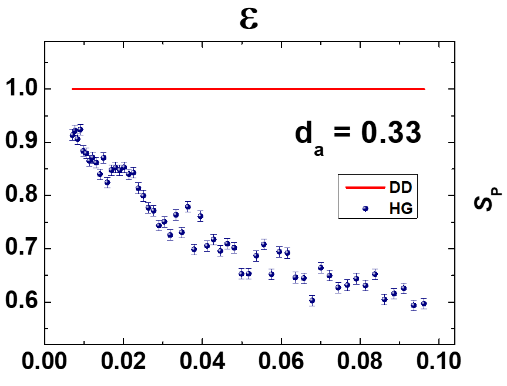}
\caption{\textbf{Scalar product quantifier.} 
The data points show the experimentally estimated scalar product in Eq.~(\ref{Spdef}), plotted as function of $\epsilon$ at fixed  $d_a$; 
$\mathcal{S}_P(d_a=0.33,\epsilon)$
The error bars show our estimate of the statistical errors.
At $\epsilon=0$ the scalar product is lower than unity (about $0.92$) because of systematic uncertainty not considered in error bars. 
By considering both statistic and systematic uncertainties we have an overall uncertainty of about $0.09$.
We argue that when the HG scalar product is below $0.91$ (when $\epsilon$ is larger than $0.008$), the SPADE technique can distinctly discriminate between star and exoplanet.
The solid line shows the analogous quantity for DD, which is close to one across all the range up to $\epsilon=0.1$.
}
\label{fi:spe}
\end{figure}

Figure~\ref{fi:spd} and~\ref{fi:spe} show the quantity 
$\mathcal{S}_P(d_a,\epsilon)$ 
as a function of $d_a$ at fixed $\epsilon$ and 
$\mathcal{S}_P(d_a,\epsilon)$ as function of $\epsilon$ at fixed $d_a$, with 
$\mathcal{S}_P(d_a,\epsilon=0.021)$ and $\mathcal{S}_P(d_a=0.33,\epsilon)$ respectively.
To compare with direct detection, we have computed the quantity $\mathcal{S}_P$ that would be obtained by implementing the spectral analysis through direct detection.
To this goal, we have simulated two direct detection systems centered with exoplanet and with star.
Using the same number of photons of the HG SPADE experiment 
($N_A\simeq 148000$) 
the direct detectors aligned with exoplanet {(at coordinate $y_1=0,y_2=w_0 d_a$ on image plane)} and star (at coordinate $y_1=0,y_2=0$ on image plane) acquired $\text{dd}_{p}$ and $\text{dd}_{s}$ photons respectively.
The two systems collect the photons over two spatial regions with radius $w_0$ centered to exoplanet and star, and spectrally disperse them with the same resolution used for HG spade $\simeq 1 \, nm$.
We compute
\begin{align}
    \text{dd}_{p}(\lambda)
    & = I_1  f_s(\lambda)  N_A (1-\epsilon) +  I_2 f_p(\lambda) N_A \epsilon \, , \\
    \text{dd}_{s}(\lambda)
    & = I_3  f_s(\lambda)  N_A (1-\epsilon) +  I_4 f_p(\lambda) N_A \epsilon \, ,
\end{align}
where
\begin{align}
    I_1 & = \int_{-w_0}^{w_0}\,dy_1\int_{w_0d_a - w_0}^{w_0d_a + w_0} G_s(y_1,y_2)\,dy_2 \, , \\
    I_2 & = \int_{-w_0}^{w_0}\,dy_1\int_{w_0d_a - w_0}^{w_0d_a + w_0} G_p(y_1,y_2)\,dy_2 \, , \\
    I_3 & = \int_{-w_0}^{w_0}\,dy_1\int_{w_0}^{+w_0} G_s(y_1,y_2)\,dy_2 \, , \\
    I_4 & = \int_{-w_0}^{w_0}\,dy_1\int_{w_0}^{+w_0} G_p(y_1,y_2)\,dy_2 \, ,
\end{align}
and 
\begin{align}
    G_s(y_1,y_2) & = \frac{1}{2 \pi w_0^2}  e^{-\frac{-y_1^2+y_2^2}{2 w_0^2}} \, , \\
    G_p(y_1,y_2) & = \frac{1}{2 \pi w_0^2}  e^{-\frac{-y_1^2+(y_2-d_a w_0)^2}{2 w_0^2}} \, .
\end{align}

In analogy with Eqs.~(\ref{CNeq1})-(\ref{CNeq2}), we define the normalized vectors
\begin{align}
\text{DD}_p(\lambda) & := \frac{ \text{dd}_p(\lambda) }{\sqrt{ \sum_{\lambda'} \text{dd}_p(\lambda')^2 }} \, , \\
\text{DD}_s(\lambda) & := \frac{\text{dd}_s(\lambda)}{ \sqrt{ \sum_{\lambda'} \text{dd}_s(\lambda')^2}} \, .
\end{align}

Considering the same number of photons, the uncertainty is of the order of a few per cent and, as shown in the figures ~\ref{fi:spd} and~\ref{fi:spe}, the scalar product  $\mathcal{S}_P ^\text{DD}=\sum_{\lambda_i} \text{DD}_{s}(\lambda_i) \text{DD}_{p}(\lambda_i)$ for direct detection spectroscopy remains close to one in the region of interest and is much larger than scalar product $\mathcal{S}_P$ in HG SPADE case. 
This corresponds to the fact in the sub-Rayleigh regime ($d_a < 1$) HG SPADE is more capable that direct detection to distinguish the photons coming from the secondary source, hence allowing a better estimation of their spectral properties.



%

\section{Conclusions}\label{sec:end}

In this work we demonstrate a scheme for spectroscopy with spatial super-resolution. The scheme is based on spatial-mode demultiplexing (SPADE) of the optical field in its transverse components. We make use of the demultiplexer PROTEUS-C model from Cailabs, which decomposes the transverse field along the Hermite-Gaussian (HG) components. 
HG SPADE has been widely studied in the past few years, following the seminal work of Tsang \textit{et al.}~\cite{Tsang2016}, as a means to achieve sub-Rayleigh estimation and discrimination of quantum states. Potential applications have been proposed to enhance astronomic observations, including the observation of exoplanets~\cite{PhysRevA.96.062107,PhysRevLett.127.130502,GraceGuha}.
{SPADE-based super-resolution imaging has been implemented in several ways, including spatial light modulators~\cite{paur2016achieving,Paur:18,Zhou2019,Lvovsky,Zhou:23,Frank:23}, 
image inversion~\cite{Tham2017}, 
photonic lantern~\cite{Salit:20}, and multi-plane light conversion~\cite{Boucher2020,Santamaria22,Treps23,Amato23}, but but so far none of these has ever been applied to spectroscopy.}
Here we demonstrate the first application of HG SPADE to achieve super-resolved spectroscopy. 

One of the challenges addressed by exoplanet science is to determine the atmospheric makeup of exoplanets. 
In particular, one searches for biomarkers as oxygen or methane, whose presence is witnessed by  absorption lines in the visible and near-infrared spectrum~\cite{bio1,bio2}. 
Gaining information about the spectrum of the exoplanet is hard because of the angular vicinity to the star. Most of the photons collected come directly from the star, and it is hard to discriminate the relatively few photons that contain information about the planetary atmosphere.
In our proof-of-principle experiment we simulate the observation and spectral analysis of a system composed of a primary source (the star) and a secondary source (the planet). 

In principle, HG SPADE allows us to completely decouple the photons coming from the secondary source. Due to symmetry, if the demultiplexer is aligned towards the primary source, then only the photons from the secondary source are collected into the excited mode $\text{HG}_{1}$.
In practice, the scheme is limited by cross talk due to experimental imperfections in the demultiplexer as well as residual misalignment.
Here we introduce a quantity, which can be directly measured experimentally, that quantifies the capability of the system to distinguish the photons emitted by the secondary source and estimate their spectral properties. 
The crucial parameters are the relative intensity of the planet compared to the star, and their angular separation. 
We shows that there exists a regime, where the relative intensity and the angular separation are not too small, where 
HG SPADE is capable of extracting substantial information about the spectrum of the secondary source, in particular much more than direct detection. 
These results demonstrate the potential usefulness of SPADE in exoplanet spectroscopy and
pave the way to experiments beyond proof of principle,  towards in-field demonstration of super-resolution spectroscopy from spatial-mode demultiplexing.

\vspace{0.5cm}
\noindent
\textbf{Funding.}
European Union — Next Generation EU through 
PRIN 2022 (CUP:D53D23002850006)
and
PNRR MUR projects PE0000023-NQSTI.
Italian Space Agency (ASI, Agenzia Spaziale Italiana) through projects 
‘Subdiffraction Quantum Imaging’ (SQI) n.~2023-13-HH.0 and 
‘MOlecular spectroscopy for Space science and quantum physics Test’ (MOST).\\
\textbf{Acknowledgments.}
We thank Cailabs, 38 boulevard Albert 1er, 35200 Rennes, France.\\
\textbf{Disclosures.}
The authors declare no conflicts of interest.\\
\textbf{Data Availability Statement.}
Data available from the authors on request.\\
\textbf{Authors contribution.}
L.S.A. designed the experimental setup and realized the apparatus; C.L. developed the theory and mathematical modeling, and assisted with the experimental design, and supervised the experiment; L.S.A. and F.S. performed data acquisition and analysis; D.P. developed the instruments control software; all authors discussed the results and contributed to the final manuscript.



\providecommand{\noopsort}[1]{}\providecommand{\singleletter}[1]{#1}%

\end{document}